\documentclass[12pt]{JHEP}
\usepackage{latexsym}
\usepackage{epsfig}
\usepackage{amssymb}
\usepackage{amsthm}
\usepackage{amsmath}
\usepackage{amssymb,amsfonts}
\usepackage{verbatim}

\def\be{\begin{equation}}
\def\ee{\end{equation}}
\def\ba{\begin{eqnarray}}
\def\ea{\end{eqnarray}}
\def\ben{\begin{equation*}}
\def\een{\end{equation*}}
\def\ban{\begin{eqnarray*}}
\def\ean{\end{eqnarray*}}

\def\cF{{\cal F}}

\def\cK{{\cal K}}

\def\cM{{\cal M}}
\def\cN{{\cal N}}
\def\cO{{\cal O}}

\def\cS{{\cal S}}

\def\bS{{\bf S}}
\def\bT{{\bf T}}

\def\bZ{{\bf Z}}

\def\bbR{{\mathbb R}}
\def\bbZ{{\mathbb Z}}

\def\ii{{\rm i}}

\def\Im{{\rm Im} \, }
\def\diag{{\rm diag}}

\def\nn{\nonumber}

\newcommand{\eqr}[1]{(\ref{#1})}

\title{Flat - brane compactifications in Supergravity induced by scalars}

\author{Alex Kehagias and Constantina Mattheopoulou\\
    Department of Physics, National Technical University of Athens, GR-15773, Zografou, Athens,
    Greece\\
    E-mail: \email{kehagias@central.ntua.gr, conmat@central.ntua.gr}}

\abstract{We discuss flat compactifications of supergravities in
diverse dimensions in the presence of branes. The compactification
is induced by the scalar fields of supergravity and it is such
that there is no relic cosmological constant on the brane,
rendering this way the latter flat.
We discuss in particular the $D=4$, $\cN=2,4$ and $D=8$, $\cN=1$
supergravities  with $n=1,2,3$ vector multiplets where the scalar
manifolds are Grassmannian cosets  of the form  $SO(2,n)/SO(2)
\times SO(n)$. By introducing branes at certain points in the
transverse space, finite energy solutions to the field equations
are constructed. Some of the solutions we present may be
interpreted as intersecting branes.}

\begin{document}

\section{Introduction}

Codimension-two brane solutions of gravitational theories have
attracted much interest the recent years. The distinguishing
features of such solutions are: (i) the brane worldvolume is
always Ricci-flat, irrespective of any vacuum energy induced on
the brane, and (ii) the internal space has the same local geometry
as it would have in the absence of branes, apart from
conical deficit angles proportional
to the brane tensions
\cite{Sundrum:1998ns}-\cite{Navarro:2003bf}.
The first property motivates the study of such
solutions in the context of six-dimensional theories, since
 the corresponding Ricci-flat 3-branes   provide a new
perspective for a possible solution of the cosmological constant
problem \cite{Weinberg}-\cite{Chen:2000at}. More generally, codimension-two brane solutions can be
examined in gravitational theories in diverse dimensions, where
they correspond to defects of dimension lower or higher than four.

The codimension-two solutions mentioned above may be triggered by
matter fields appearing in the theory, such as $p$-form gauge
fields \cite{Carroll,Navarro:2003vw,Peloso:2006cq,Aghababaie:2003ar} or, most importantly,  by scalar
fields. On the other hand, for the case of sigma models with a compact target
space, solutions of this type have been found in \cite{RandjbarDaemi:2004ni,Lee:2004vn};
however,  sigma models with such scalar manifolds do not
occur in supergravity. For the case of
non-compact sigma models, there are two prototype solutions. The
first type of solutions \cite{Kehagias:2004fb} generalize the
``teardrop'' solution of \cite{Gell-Mann1} to account for the
presence of branes; here the internal 2-dimensional manifold is a
{\it non-compact} space of finite volume~\cite{Nicolai,Kehagias:2000dg,Cohen:1999ia}
 and the geometry has a
naked singularity at its boundary which, however, may be rendered
harmless by imposing appropriate boundary conditions. These boundary conditions
guarantee that the conservation laws of the theory are not spoiled
 and energy, momentum angular momentum etc do not ``leak" from the boundary. The second type of solutions are
based on the ``stringy cosmic string'' of \cite{Greene:1989ya}. In
this case, the internal geometry can be non-singular provided that
the brane tensions are restricted to a certain range, and,
in fact, correspond to a compact manifold of Euler number 2
provided that the brane tensions are appropriately fine-tuned~\cite{Nair:2004yu}. Moreover, the existence of modular
symmetries in the non-compact case guarantees that the scalars and
the metric are actually single-valued, unlike the compact case where this
issue is not clear.

In this paper we present codimension-two
solutions of supergravity models in diverse dimensions, in the
presence of branes. In particular, we consider $D$-dimensional
supergravity theories coupled to nonlinear sigma models, with the
sigma-model target spaces being non-compact K\"ahler manifolds.
We seek exact solutions of the form $M^{D-n}
\times \cK$, where $M^{D-n}$ is a flat Minkowski
space and $\cK$ is an $n$-dimensional internal space. As concrete
examples, we consider the cases of $\cN=4$ and $\cN=2$
supergravity in 4 dimensions (with the solutions corresponding to
strings), as well as the cases of minimal supergravities coupled
to vector multiplets in 8 dimensions (with the solutions
corresponding to parallel or intersecting  five-branes). For all of the above cases, the scalar
 manifold is  special K\"ahler  of the form
${SL(2,\bbR)\over U(1)} \times {SO(2,n) \over {SO(2)\times
SO(n)}}$ \cite{de Wit:1984pk,Ceresole:1995jg,Ferrara:1987tp} or a
Grassmannian coset  ${SO(2,n) \over {SO(2)\times
SO(n)}}$ \cite{Salam:1985ns}. To find the explicit solutions, we
follow the guidelines of \cite{Greene:1989ya}, employing a
holomorphic ansatz for the scalars that restricts the latter to
lie in the fundamental domain of the modular groups
and allowing modular $SL(2,\bbZ), ~SL(2,\bbZ)\times SL(2,\bbZ)$ or $Sp(4,\bbZ)$  jumps around certain points in the internal space.
This leads to scalar field configurations with finite
energy per unit volume. Explicit solutions may be
presented only if $n\leqslant 3$, since it is only for these cases
that the modular forms required to construct the solutions are
explicitly known. The solutions under consideration generically
possess singularities appearing in the form of conical deficit
angles; in order to arrive at non-singular solutions, one has to
arrange the total deficit angle to be
$4 \pi$ in which case the internal space compactifies to $\bS^2$.

The solutions described above can be generalized to include  brane
configurations as well. In this case, the branes are introduced  at the points where the scalar fields  diverge
leading to extra delta-function contributions to the scalar
energy-momentum tensor. The branes induces further deficit angles,
proportional to their tensions in the internal space. In such
a scenario, the requirement for the absence of conical
singularities may be fulfilled by suitably
tuning\cite{Sundrum:1998ns,RandjbarDaemi:2004ni,Nair:2004yu} the
brane tensions  so that the total deficit angle equals $4 \pi$.

The structure of this paper is as follows. In section 2, we review
the solution of the equations of motion for a gravitational theory
coupled to a K\"ahler sigma model, we examine the prototype
stringy cosmic string solution, and we also comment of the issue
of codimension-four solutions. In section 3, we describe the
essential aspects of the K\"ahler sigma models under consideration
and we state the explicit form of the K\"ahler potentials in terms
of the supergravity fields. In section 4, we apply the above
results to the cases $\cN=4$ and $\cN=2$ supergravity in 4
dimensions and we present the corresponding ``stringy cosmic
string'' solutions, while we also consider the case of minimal
supergravity in 8 dimensions and we present the corresponding
5-brane solution as well as a four-dimensional intersecting-brane
solution. Finally, in section 5 we summarize our main results.

\section{K\"ahler sigma models}

Our general setup corresponds to a $D$-dimensional theory of
gravity coupled to a K\"ahler sigma model. This is a generic
situation in almost all supergravity theories in four or higher
dimensions. Without specifying the scalar sigma model in detail at
this stage, we assume that its target space is a K\"ahler manifold
$\cM$ spanned by the complex coordinates $(\varphi^i ,
\bar{\varphi}^{\bar{j}})$ and characterized by the K\"ahler
potential $K(\varphi^i,\bar{\varphi}^{\bar{j}})$ and the metric
$K_{i \bar{j}}(\varphi^i,\bar{\varphi}^{\bar{i}}) =
\partial_i \partial_{\bar{j}} K(\varphi^i,\bar{\varphi}^{\bar{j}})$. The dynamics of this system is described by the
action
\be
\label{2-1} S = \int d^D x \sqrt{-g} M_*^{D-2} \left( {1 \over 2}
R - K_{i\bar{j}}(\varphi^i,\bar{\varphi}^{\bar{j}}) \partial_M
\varphi^i \partial^M \bar{\varphi}^{\bar{j}} \right)\ ,
\ee
where $M_*^{D-2}$ is the $D-$dimensional Planck
mass.
The equations of motion as follow from (\ref{2-1}) are the scalar
equation of motion
\be
\label{2-2} {1 \over \sqrt{-g}} \partial_M \left( \sqrt{-g}
K_{i\bar{j}} \partial^M \varphi^i \right) - \partial_{\bar{j}}
K_{i
\bar{k}} \partial_M \varphi^i \partial^M \bar{\varphi}^{\bar{k}} = 0\ ,
\ee
and the Einstein equation
\be
\label{2-3} R_{MN} = K_{i \bar{j}} (\partial_M \varphi^i
\partial_N
\bar{\varphi}^{\bar{j}} + \partial_M \bar{\varphi}^{\bar{j}}
\partial_N \varphi^i)\ .
\ee
We are looking for solutions of the form $M^{D-n}\times \cK$,
where $M^{D-n}$ is a flat $(D-n)$-dimensional Minkowski space-time
parametrized by the coordinates $x^\mu$ and $\cK$ is an internal
complex manifold, parametrized by the complex coordinates
$(z^a,\bar{z}^{\bar{b}})$ and metric
$k_{a\bar{b}}(z^a,\bar{z}^{\bar{b}})$. To solve
\eqr{2-2},\eqr{2-3}, we assume that the scalars depend only on the
internal coordinates, so that the solution we are after is
\be
\label{2-4} ds^2 = \eta_{\mu\nu} d x^\mu d x^\nu +
k_{a\bar{b}}(z^a,\bar{z}^{\bar{b}}) dz^a d\bar{z}^{\bar{b}}\ ,~~~~
\varphi^i = \varphi^i (z^a,\bar{z}^{\bar{b}})\ .
\ee
Inserting this ansatz in Eqs. \eqr{2-2} and \eqr{2-3}, and making
use of standard K\"ahler identities such as $\partial_{[k}
K_{i]\bar{j}} = 0$ and $\partial_{[c} k_{a]\bar{b}} = 0$, we find
that these equations reduce to
\be
\label{2-5} k^{a\bar{b}} \partial_a \partial_{\bar{b}} \varphi^i +
K^{i \bar{j}} \partial_k K_{l \bar{j}} k^{a\bar{b}} \partial_a
\varphi^k \partial_{\bar{b}} \varphi^l = 0\ ,
\ee
and
\be
\label{2-6} \partial_a \partial_{\bar{b}} \ln \det k_{a \bar{b}} =
- K_{i \bar{j}} (\partial_a \varphi^i \partial_{\bar{b}}
\bar{\varphi}^{\bar{j}} +
\partial_a \bar{\varphi}^{\bar{j}} \partial_{\bar{b}} \varphi^i)\ ,
\ee
respectively. Starting from the scalar equation \eqr{2-5}, we see
that it is automatically satisfied when the $\varphi^i$ are
holomorphic or antiholomorphic functions. Restricting for
definiteness to the holomorphic case,
\be
\label{2-7} \varphi^i=\varphi^i(z^a)\ ,
\ee
we find that the Einstein equation \eqr{2-6} reduces to
\be
\label{2-8} \partial_a \partial_{\bar{b}} \ln \det k_{a \bar{b}} =
- K_{i \bar{j}} \partial_a \varphi^i \partial_{\bar{b}}
\bar{\varphi}^{\bar{j}} = - \partial_a \partial_{\bar{b}} K\ ,
\ee
and hence it is solved by
\be
\label{2-9} \det k_{a \bar{b}}(z^a,\bar{z}^{\bar{b}}) = e^{-
K(\varphi^i,\bar{\varphi}^{\bar{j}})} |F(z^a)|^2\ ,
\ee
where $F(z^a)$ is an arbitrary holomorphic function. Given a
specific scalar manifold $\cM$ and a specific ansatz for the
moduli $\varphi^i(z^a)$, the choice of $F(z^a)$ is dictated by
the symmetries of the moduli space and by geometric  properties such as the  non-degeneracy of the metric
and the absence of curvature
singularities.

Generalizing our solution, we may also consider a $(D-n-1)$-brane
located at $z^a=0$ to which the scalar fields are not coupled.
These branes contribute an additional energy momentum tensor to
the right-hand side of Einstein equations of the
form\footnote{where $\int d^n z \sqrt{\det k_{a
\bar{b}}}\delta^{(n)}(z)=1$}
\be
\label{2-10}T_{\mu\nu}=-{g_{\mu\nu}} T_0 \delta^{(n)}(z) \,
~~~T_{ab}=T_{a
\bar{b}}=T_{\bar{a} \bar{b}}=0
\ee
where $\mu,\nu = 0,\ldots,D-n-1$ and $T_0$ is the tension of the
brane located at the origin. Then, \eqr{2-8} is changed to
\be
\label{2-11} \partial_a \partial_{\bar{b}} \ln \det k_{a \bar{b}}
= - K_{i \bar{j}} \partial_a \varphi^i \partial_{\bar{b}}
\bar{\varphi}^{\bar{j}} -{k_{a \bar{b}}} \frac{T_0}{M_*^{D-2}} \delta^{(n)}(z) ,
\ee
which can be solved under certain conditions, as we will see
below.

\subsection{Complex dimension one}

We will first consider the case where the internal space has
complex dimension one, which is relevant for seeking
codimension-two brane solutions with an internal compact or
non-compact space. Parametrizing the transverse space by the
complex coordinate $z$, the explicit form of the solution
\eqr{2-9} reads
\be
\label{2-12} \varphi^i = \varphi^i(z)\ , \qquad ds^2 =
\eta_{\mu\nu} d x^\mu d x^\nu + e^{-K} |F(z)|^2 dz d\bar{z}\ \ .
\ee
The energy density of such configurations can be written in the
BPS-like form
\be
\label{2-13} E = {\rm i \over 2} \int_{\cK} d^2 z K_{i\bar{j}}
\partial \varphi^i \bar{\partial} \varphi^{\bar{j}} = {\rm i \over
2} \int_{\varphi(\cK)} \partial
\bar{\partial} K\ ,
\ee
where $\partial$ and $\bar{\partial}$ are Dolbeault operators and
in the second integral, the domain of integration has been pulled
back to the image $\varphi(\cK)$ of the internal manifold $\cK$.
In the presence of brane with tension $T_0$, the above relations
are modified to
\be
\label{2-14} ds^2 = \eta_{\mu\nu} d x^\mu d x^\nu + e^{-K}
|F(z)|^2 |z|^{-T_0/{\pi M_*^{D-2}}} dz d\bar{z}
\ee
and
\be
\label{2-15} E = {\rm i \over 2} \int_{\cK} d^2 z K_{i\bar{j}}
\partial \varphi^i \bar{\partial} \varphi^{\bar{j}} + \frac{T_0}{M_*^{D-2}} \ .
\ee
Although the first form of the integral in \eqr{2-13} may appear to give a zero
answer due to the fact that the integration domain is a compact
surface, the K\"ahler potential is not globally well-defined on
$\cM$ and hence the integral may contain jumps that render it
nonzero. This is made more explicit in the second form of the
integral in terms of the K\"ahler potential. Indeed, if there exist symmetries of $\cM$ (such as
modular invariance for example) that result in a bounded $\varphi(\cK)$  of finite volume, the integral may give a nonzero
result due to boundary terms. The energy per unit volume is
alternatively expressed as
\be \label{2-151}
 E = 2 \pi \chi\ ,
\ee
where $\chi$ is the Euler characteristic of $\cK$.

\subsubsection{Stringy cosmic strings}

An example of particular importance, which will serve as the
prototype of the solutions to be constructed later on, refers to
the case where the moduli space consists of a single toroidal
modulus $\tau=\tau_1+{\rm i}\tau_2$, such as the type IIB
axion-dilaton or the complex-structure or K\"ahler modulus of an
internal compactification torus. In this case, the scalar manifold
is the $SL(2,\mathbb{R})/U(1)$ space, which is a K\"ahler manifold
with K\"ahler potential
\be
\label{2-16} K = - \ln \left( \ii (\tau -\bar{\tau}) \right)\ ,
\ee
so that the effective action \eqr{2-1} takes the explicit form
\be
\label{2-17} S = \int d^D x \sqrt{-g} M_*^{D-2}\left( {1 \over 2} R +
{\partial_\mu \tau
\partial^\mu \bar{\tau} \over (\tau-\bar{\tau})^2}\right)\ .
\ee
The solution to the equations of motion for the above action as we
have seen are
\be
\label{2-18} \tau=\tau(z)\ ,\qquad ds^2 = \eta_{\mu\nu} d x^\mu d
x^\nu + \tau_2 | F(z)|^2 dz d\bar{z}\ ,
\ee
whose energy per unit volume reads
\be
\label{2-19} E = - {\rm i \over 2} \int_{\cK} d^2 z \partial
\bar{\partial} \ln \tau_2\ .
\ee

To discuss the possible choices of $\tau(z)$, we first note that
the effective action \eqr{2-15} has a symmetry under the
$SL(2,\bbR)$ group, acting as
\be
\label{2-20} \tau \to {a \tau+b \over c\tau + d}\ , \qquad
ad-bc=1\ .
\ee
In the full theory, this symmetry group is broken to the modular
group $PSL(2,\bbZ)$, which is interpreted as a local symmetry. As
a result, the space of inequivalent choices of the modulus $\tau$
is the quotient of the complex $\tau$ plane by the $PSL(2,\bbZ)$
group, which can be taken to be the fundamental domain $\cF_1$
specified by the conditions
\be
\label{2-21} |\tau_1| \leqslant {1 \over 2}\ ,\qquad \tau_2 > 0\
,\qquad |\tau| \geqslant 1\ ,\qquad \tau_1 \leqslant 0 \hbox{ if }
|\tau|=1\ .
\ee
The above considerations imply that an arbitrary holomorphic
ansatz for $\tau(z)$ is generally inconsistent since $\tau$ is
restricted to live on $\cF_1$ while $z$ covers the whole Riemann
sphere; this is alternatively verified by noting that for such a
naive choice, e.g. $\tau(z) = z^n$, the energy per unit volume
diverges. To construct consistent, finite-energy solutions we need
a holomorphic function that provides a one-to-one mapping
fundamental domain $\cF_1$ to the Riemman sphere. This mapping is
provided by the modular function $j(\tau)$ or, equivalently, by
Klein's absolute invariant
\be
\label{2-22} J(\tau) = j(\tau) - 744 = {{41 {E_4(\tau)}^{3} + 31
{E_6(\tau)}^{2}} \over {72 \Delta(\tau)}}\ , \ee where
$E_4(\tau)$, $E_6(\tau)$ are the Eisenstein series of weight $4$
and $6$ respectively, and $\Delta(\tau) = ( {E_4(\tau)}^{3} -
{E_6(\tau)}^{2}) / 1728$ is the cusp form of weight twelve. The
modular invariant has the asymptotic behavior
\be
\label{2-23} J(\tau) \sim q^{-1}\ ,\qquad \hbox{ as }
\tau_2 \to \infty\ . \ee
Finite-energy solutions can then be constructed by equating
$J(\tau(z))$ to a holomorphic function of $z$. Doing so, pulling
back the integral in \eqr{2-19} from the $z$-plane to $\cF_1$, and
converting it into a line integral over the boundary of $\cF_1$,
we indeed obtain the finite expression
\be
\label{2-24} E = - {{\rm i} N \over 2} \int_{\cF_1} d^2\tau
\partial_\tau
\partial_{\bar{\tau}} \ln \tau_2\ = {\pi \over 6} N\ ,
\ee
where $N$ is the number of times the $z$-plane covers $\cF_1$.
Here, we will consider the rational maps
\be
\label{2-25} J(\tau(z)) = {P(z) \over Q(z)}\ ,
\ee
where $P(z)$ and $Q(z)$ are polynomials in $z$ of degrees $p$ and
$q$ respectively. For this choice, the integer $N$ is equal to $q$
if $p \leqslant q$ (in which case $J(\tau)$ approaches a constant
value at infinity and diverges as $(z-z_i)^{-1}$ at the zeros of
$Q(z)$ which are identified with the ``cores'' of the solutions)
and equal to $p$ if $p>q$ (in which case $J(\tau)$ diverges as
$z^{p-q}$ at infinity). In what follows we will consider the $p <
q$ case where $J=b/{(z-z_i)}$. Given this ansatz for $\tau(z)$, it
remains to choose the function $F(z)$ in \eqr{2-18} in such a way
that the metric is modular-invariant and non-degenerate. The first
requirement is fulfilled by noting that the $PSL(2,\bbZ)$
transformation of $\tau_2$ is given by
\be
\label{2-26} \tau_2 \to {\tau_2 \over | c \tau + d |^2}\ ,
\ee
and hence can be compensated by multiplying $\tau_2$ by
$|f(\tau)|^2$ where $f(\tau)$ is an $PSL(2,\bbZ)$ modular form of
weight $1$. This modular form is explicitly given in terms of the
square of Dedekind eta function
\be
\label{2-27} \eta(\tau) = q^{1/24} \prod_{n=1}^\infty (1-q^n) =
\Delta^{1/24}\ ,\qquad q=e^{2\pi {\rm i}\tau} .
\ee
So the combination $\tau_2 |\eta(\tau)|^4$ is modular invariant.
To fulfil the second requirement we note that, near the zeros
$z_i$, we can write $q^{-1} \sim J(\tau) \sim (z-z_i)^{-1}$ which
implies from \eqr{2-27} that $\tau_2 |\eta(\tau)|^4 \sim
|(z-z_i)^{1/12}|^2$. Therefore, the choice $F(z)=\eta(\tau)^2
\prod_{i=1}^N (z-z_i)^{-1/12}$ leads to the modular-invariant,
non-degenerate solution
\be
\label{2-28} ds^2 = \eta_{\mu\nu} d x^\mu d x^\nu + \tau_2
|\eta(\tau)|^4 \left|\prod_{i=1}^N (z-z_i)^{-1/12}\right|^2 dz
d\bar{z}\ ,\qquad J(\tau(z)) = {P(z) \over Q(z)}\ .
\ee
As $|z| \to \infty$, the internal metric approaches $k_{z\bar{z}} \sim
(z\bar{z})^{-N/12}$. Hence, by standard arguments, the solution
has a deficit angle $\delta = {\pi \over 6} N$, which is equal to
the energy as expected.  So, at infinity, the internal space has
conical singularities which signify the geometry of a non-compact
space. For $N=12$ the internal space is cylindrical, while for
$N=24$ the internal space compactifies to $\bS^2$.

The above considerations are quite general and apply in all cases
where toroidal moduli with $PSL(2,\bbZ)$ symmetry exist.  For
instance, we can consider the case of $D=4$ theories arising e.g.
from string compactifications on an internal space containing a
$\bT^2$. Then the resulting solution \eqr{2-28} corresponds to a
configuration of $N$ strings carrying charge under the toroidal
modulus field. This solution is the stringy cosmic string of
\cite{Greene:1989ya}.

There is also the
possibility to add $(D-3)-$ brane sources at the points $z_i$
where the scalar field diverges.
In this case,  an energy-momentum tensor of the form
\be
\label{2-29} T_{\mu\nu}=-{\eta_{\mu\nu}} \sum_{i=1}^N {T_i}
\delta^{(2)}(z-z_i) \, ,~~~~~T_{z\bar{z}}=0\, ,
\ee
should be introduced, where $T_i$ is the tension of the brane located at the point
$z_i$. These branes cause additional deficit angles
equal to their tensions. Then, from \eqr{2-14} we find that the
modular-invariant metric is
\be
\label{2-30} ds^2 = \eta_{\mu\nu} d x^\mu d x^\nu + \tau_2
|\eta(\tau)|^4 \left|\prod_{i=1}^N (z-z_i)^{-1/12}\right|^2
\left|\prod_{i=1}^N (z-z_i)^{-{T_i}/{2 \pi {M_*^{D-2}}}}\right|^2
dz d\bar{z}\ .
\ee
At infinity the total deficit angle turns out to be
\be
\label{2-31} \delta=\frac{\pi}{6} N+\sum_{i=1}^N
\frac{T_i}{M_*^{D-2}}\ ,
\ee
which is equal to the energy \eqr{2-151} as expected. In order for
the internal space to compactify to $\bS^2$, the above deficit
angle must be equal to $4\pi$. This amounts to a fine-tuning
condition on the brane tensions $T_i$, namely $\sum_i^N {T_i}=4\pi
{M_*^{D-2}}(1- {N \over 24})$.

At the vicinity of each brane $(z\to z_i)$, where
$J(\tau)\to\infty$, the internal metric becomes $k_{z\bar{z}}\sim
\tau_2(z-z_i)^{-{T_i}/{\pi {M_*^{D-2}}}}$. Then, contracting
\eqr{2-11} with $k^{z\bar{z}}$ one deduces that the Ricci scalar
is not singular for $T_i > {2 \pi M_*^{D-2}}$. But the true
condition for the absence of curvature singularities follows when
the previous condition and Eq. \eqr{2-31} are both satisfied.
These conditions restrict the number of branes to $N=1$, in
accordance with the result of \cite{Nair:2004yu}. Note that in the
absence of the extra term involving the tension in $k_{z\bar{z}}$
there is a curvature singularity.

\subsection{Complex dimension two}

We next proceed to the case where the internal space has complex
dimension two \cite{Hung:2006jh}, which is relevant for seeking
codimension-four brane solutions with an internal compact or
non-compact space. Now, Eq. (\ref{2-9}) takes the form
\be
\label{2-32} k_{1 \bar{1}} k_{2 \bar{2}} - k_{1 \bar{2}} k_{2
\bar{1}} = e^{-K} |F(z,w)|^2\ ,
\ee
which is a highly nonlinear differential equation. Equation
\eqr{2-32} is very difficult to be solved given an explicit form
of the K\"ahler potential $K$ of the scalar manifold and a general
holomorphic ansatz for the fields $\varphi^i(z,w)$. However, for
the special case where $K$ decomposes as the sum
\be
\label{2-33} K(\varphi^i,\bar{\varphi}^{\bar{j}}) = K^{(1)}
(\varphi^A,\bar{\varphi}^{\bar{A}}) + K^{(2)}
(\varphi^B,\bar{\varphi}^{\bar{B}})\ ,
\ee
with the first term involving a subset
$(\varphi^A,\bar{\varphi}^{\bar{A}})$ of the
$(\varphi^i,\bar{\varphi}^{\bar{i}})$ and the second term
involving the remaining fields
$(\varphi^B,\bar{\varphi}^{\bar{B}})$, we can easily solve this
equation by assuming an ansatz of the form
\be
\label{2-34} \varphi^A = \varphi^A(z)\ ,\quad \varphi^B =
\varphi^B(w)\ ,\qquad k = k^{(1)} (z,\bar{z}) + k^{(2)}
(w,\bar{w}) \ ,
\ee
where the $\varphi^A$ and $\varphi^B$ depend only on $z$ and $w$
and the metric is the sum of two terms involving $(z,\bar{z})$ and
$(w,\bar{w})$ respectively. Writing also $F(z,w) = F^{(1)}(z)
F^{(2)}(w)$, Eq. \eqr{2-32} simplifies to
\be
\label{2-35} k^{(1)}_{1 \bar{1}} k^{(2)}_{2 \bar{2}} = \left[
e^{-K^{(1)}} |F^{(1)}(z)|^2 \right] \left[ e^{-K^{(2)}}
|F^{(2)}(w)|^2 \right]\ ,
\ee
and is easily solved by taking $k^{(1)}_{1 \bar{1}}$ and
$k^{(2)}_{2 \bar{2}}$ equal to the first and second terms in
brackets respectively. The final solution, which generalizes
\eqr{2-30} then reads
\ba
\label{2-36} &\varphi^A = \varphi^A(z)\ ,\qquad \varphi^B =
\varphi^B(w)\
,\nn\\
& ds^2 = \eta_{\mu\nu} d x^\mu d x^\nu + e^{-K^{(1)}}
|F^{(1)}(z)|^2 \left|\prod_{i=1}^{N_1} (z-z_i)^{-{T^{(1)}_i}/{2
\pi {M_*^{D-2}}}}\right|^2
dz d\bar{z} +\nn \\
& e^{-K^{(2)}} |F^{(2)}(w)|^2 \left|\prod_{j=1}^{N_2}
(w-w_j)^{-{T^{(2)}_j}/{2 \pi {M_*^{D-2}}}}\right|^2dw d\bar{w}\ .
\ea
We will discuss in section 4.3 the interpretation of such a solution.

\section{Special K\"ahler and Grassmannian }

The construction of solutions of the type described in the
previous section carries over to more complicated moduli spaces.
Here, we will construct solutions of this form for the cases where
the classical moduli space is a special K\"ahler manifold of the
form
\be
\label{3-1a} \cS\cK_{n+1} = {SL(2,\bbR) \over U(1)} \times
{SO(2,n)\over {SO(2)\times SO(n)}}\ ,
\ee
or a K\"ahler manifold of the form
\be
\label{3-1b} \cK_n = {SO(2,n)\over {SO(2)\times SO(n)}}\ .
\ee
In what follows, we will give a brief description of the geometry
of these manifolds, using the formalism of special geometry, and
we will state the corresponding K\"ahler potentials for the cases
of interest.

The geometry of the special K\"ahler manifold $\cS\cK_{n+1}$ is
completely specified by a holomorphic symplectic section \cite{de Wit:1984pk,Ceresole:1995jg}
\be
\label{3-2}\Omega = \left( \begin{array}{c} X^I \\ F_I \end{array}
\right)\ ,
\ee
in terms of which the K\"ahler potential is given by
\be
\label{3-3} K = - \ln \left( {\rm i} \langle \Omega| \bar{\Omega}
\rangle \right) \equiv - \ln \left( {\rm i} ( \bar{X}^I F_I - X^I
\bar{F}_I )\right)\ .
\ee
In the above, $X^I$, $I=0,\ldots,n+1$ are a set of complex
parameters, while $F_I$ are usually specified as the derivatives
of a holomorphic prepotential $F(X)$ with respect to the $X^I$. In
the present case, it is convenient to employ the so-called
symplectic gauge in which $\Omega$ is written as
\be
\label{3-4}\Omega =\left( \begin{array}{c} X^I \\ F_I \end{array}
\right) = \left( \begin{array}{c} X^I \\ S \eta _{IJ} X^J
\end{array} \right) \ ,
\ee
where $\eta_{IJ} = \diag(+1,+1,-1,\ldots,-1)$ is the $SO(2,n)$
invariant metric and $S$ parametrizes the ${SL(2,\bbR) \over U(1)}$
factor in the usual way.  $X^I$ parametrize the
${SO(2,n)\over {SO(2)\times SO(n)}}$ factor and are required to
satisfy the $SO(2,n)$ orthogonality condition
\be
\label{3-5}\eta_{IJ} X^I X^J = 0\ .
\ee
Although this gauge choice makes it impossible to specify $F_I$ by
means of a prepotential, Eq. \eqr{3-3} for the K\"ahler potential
is perfectly valid, leading to the result
\cite{Andrianopoli:1996cm}
\be
\label{3-6} K = K_1 + K_2\ ,
\ee
where
\be
\label{3-7} K_1 = - \ln (S - \bar{S})
\ee
is the standard K\"ahler potential for ${SL(2,\bbR) \over U(1)}$
and
\be
\label{3-8} K_2 = - \ln ( \eta_{IJ} \bar{X}^I X^J )
\ee
is the K\"ahler potential of $SO(2,n)\over {SO(2)\times SO(n)}$.
The latter
 can be verified by parametrizing $X^I$ in terms of the
independent Calabi-Vesentini coordinates $y^a$, $a=1,\ldots,n$,
according to
\be
\label{3-9} X^I(y) = \left( \begin{array}{c} {1 \over 2} (1 + y^2) \\
{{\rm i} \over 2} (1 - y^2) \\ y^a \end{array} \right)\ .
\ee
Then, it is straightforward to see that the familiar formula
\be
\label{3-10} K_2 = - \ln \left( 1 - 2 y^\dag y + |y^2|^2 \right)\,
,
\ee
is recovered.

We are particular interested in K\"ahler manifolds of the form
\eqr{3-1b} with $n=1,2,3$ for which the modular forms required to
construct our solutions are explicitly known. In what follows, we
give the explicit parametrizations of $X^I$ in terms of
supergravity fields and we state the corresponding K\"ahler
potentials for these particular cases.

\begin{itemize}
\item $n=1$. For this case, the $SO(2,1;\bbR)$ vector $X^I$ is
parametrized in terms of a single complex field $T$ as
\cite{Ceresole:1995jg}
\be
\label{3-11} X^I(T) = \left( \begin{array}{c} {1 \over \sqrt{2}} (1 - T^2) \\
- \sqrt{2} T \\ - {1 \over \sqrt{2}} (1 + T^2) \end{array}
\right)\ .
\ee
Inserting this into \eqr{3-8}, we find the K\"ahler potential
\be
\label{3-12} K_2(T) = - 2 \ln (T - \bar{T})\ .
\ee
\item $n=2$. For this case, the $SO(2,2)$ vector $X^I$ is
parametrized in terms of two complex fields $T$ and $U$
as\cite{Ceresole:1995jg,Angelantonj:2003zx}
\be
\label{3-13} X^I(T,U) = \left( \begin{array}{c} {1 \over \sqrt{2}} (1 - T U) \\
-{1 \over \sqrt{2}} (T + U) \\ -{1 \over \sqrt{2}} (1 + T U) \\ {1
\over \sqrt{2}} (T - U) \end{array} \right)\ ,
\ee
and the K\"ahler potential reads
\be
\label{3-14} K_2(T,U) = -\ln\left( (T - \bar{T}) (U - \bar{U})
\right)\ .
\ee
\item $n=3$. Now, the $SO(2,3)$ vector $X^I$ can be parametrized
in terms of three complex fields $T$, $U$ and $V$ as
\be
\label{3-15} X^I(T,U,V) = \left( \begin{array}{c} {1 \over
\sqrt{2}} ( 1 - T U + V^2 ) \\ - {1 \over \sqrt{2}}(T + U) \\ -{1
\over {\sqrt{2}}}
( 1 + T U  - V^2 ) \\ {1 \over \sqrt{2}} (T - U) \\
\sqrt{2} V
\end{array} \right)\ ,
\ee
and the K\"ahler potential reads
\be
\label{3-16} K_2(T,U,V) = - \ln\left( (T - \bar{T})(U - \bar{U}) -
(V -
\bar{V})^2 \right)\ .
\ee
\end{itemize}
To summarize, the K\"ahler potential for the special K\"ahler
manifolds \eqr{3-1a} is given by $K=K_1+K_2$ where $K_1$ is given
in \eqr{3-7} and $K_2$ is given in \eqr{3-8}, while the K\"ahler
potential for the K\"ahler manifolds \eqr{3-1b} is simply $K=K_2$.
Explicit expressions for $K_2$ for the cases $n=1$, $n=2$ and
$n=3$ are given in Eqs. \eqr{3-12}, \eqr{3-14} and \eqr{3-16}
respectively.

\section{Application to supergravity theories}

We may apply the results of the previous sections to
construct  solutions in the context of supergravity
theories where scalar K\"ahler manifolds of the sort discussed
earlier appear. In particular, we will discuss two classes of
solutions. The first class corresponds to stringy-cosmic-string
solutions of $D=4$ supergravities with $\cN=4$ or $\cN=2$
supersymmetry, arising from appropriate
heterotic string compactifications. The theories under
consideration possess modular symmetries that may be exploited to
construct stringy cosmic string solutions according to the
guidelines of section 2. Moreover, for these theories, the quantum
corrections to the K\"ahler potential are under control and thus
one can extend the classical solutions to solutions that are exact
to all orders in perturbation theory. The second class of
solutions corresponds to five-brane solutions of minimal $D=8$,
$\cN=1$ supergravity and four-dimensional intersections thereof.

\subsection{String solutions in $D=4$, $\cN=4$ supergravity}

We first consider the case of the $\cN=4$ theories
\cite{Ferrara:1987tp} arising from compactifications of the $E_8
\times E_8$ or $SO(32)$ heterotic string theories on $\bT^4 \times
\bT^2$ or, equivalently, from compactifications of $\cN=1$
six-dimensional supergravity on $\bT^2$. For these models, the
moduli space consists of three factors involving (i) the
axion-dilaton $S$, (ii) the moduli $T$ and $U$ corresponding to
the complex and K\"ahler structure moduli of $\bT^2$, and (iii)
the moduli of $\bT^4$. In what follows, we will consider only the
first two types of moduli, which parametrize the space
\be
\label{4-1} \cM = \left(PSL(2,\bbZ) \setminus {SL(2,\bbR) \over
U(1)} \right)_S \times \left( SO(2,2;\bbZ) \setminus {SO(2,2)\over
{SO(2)\times SO(2)}} \right)_{T,U}\ ,
\ee
with the isomorphism $SO(2,2;\bbZ) \cong PSL(2,\bbZ) \times
PSL(2,\bbZ)$ implying that the duality group is given by the
product $PSL(2,\bbZ)_S \times PSL(2,\bbZ)_T \times PSL(2,\bbZ)_U$.
The moduli $S$, $T$ and $U$ are given in terms of six-dimensional
fields as
\be
\label{4-2} S = \alpha + {\rm i} e^{-2\phi}\ ,\qquad T=B_{45} +
{\rm i} \sqrt{detg_{mn}}\ ,\qquad U={g_{45} \over g_{55}} + {\rm
i} {\sqrt{detg_{mn}} \over g_{55}}\ ,
\ee
where $\phi$ and $\alpha$ are the dilaton and axion while $g_{mn}$
and $B_{45}$ is the metric and $B$-field on $\bT^2$. The effective
action for these fields follows from the K\"ahler potential in Eq.
\eqr{3-14}, namely
\be
\label{4-3} K = - \ln (S-\bar{S}) - \ln (T-\bar{T}) - \ln
(U-\bar{U})\ .
\ee
It is invariant under the duality group, as well as under
string/string/string triality \cite{Duff:1995sm} which
interchanges $S$, $T$ and $U$.

The solution for the moduli for this case is readily obtained by
taking $S$, $T$ and $U$ to be holomorphic functions restricted to
the fundamental domains of $PSL(2,\bbZ)_S$, $PSL(2,\bbZ)_T$ and
$PSL(2,\bbZ)_U$, respectively, by relations of the form
\eqr{2-25}, and by inserting the K\"ahler potential \eqr{4-3} in
Eq. \eqr{2-12}. This leads to a stringy cosmic string solution
with transverse metric
\be
\label{4-4} d\sigma^2 = S_2 T_2 U_2 | F(z)|^2 dz d\bar{z}\ .
\ee
To determine $F(z)$, we impose the requirements of modular
invariance and non-degeneracy of the metric as before. The first
requirement leads to a factor of $|\eta(S) \eta(T) \eta(U)|^4$
while the second requirement leads to a factor of
$|(z-z_i)^{-1/12}|^2$ for each string. Letting $N_S$, $N_T$ and
$N_U$ be the number of strings carrying charge with respect to the
$S$, $T$ and $U$ moduli respectively, we finally find
\be
\label{4-5} d\sigma^2 = S_2 T_2 U_2 |\eta(S) \eta(T) \eta(U)|^4
\left| \prod_{i=1}^{N_S} \prod_{j=1}^{N_T} \prod_{k=1}^{N_U}
\left( (z-z_i) (z-z_j) (z-z_k) \right)^{- 1 / 12} \right|^2 dz
d\bar{z}\ .
\ee
Imposing  string/string/string triality leads to
$N_S=N_T=N_U=N$. As $|z| \to \infty$ for each string we
have a deficit angle $\delta ={\pi \over 6}$, and the energy of
the solution is $E={\pi \over 6}(N_S+N_T+N_U)={\pi \over 2} N$.
Therefore, to compactify the transverse space to $\bS^2$, we need
$N=8$.

In the above we have assumed that each string is charged with
respect to only a single modulus so $z_i\neq z_j\neq z_k$.
However, string/string/string triality also allows us to consider
``$STU$-strings'' that are charged under all three moduli. Such
configurations may give rise to orbifold singularities on the
transverse space; in order for this to occur, we need deficit
angles of the form $\delta=2\pi (n-1)/n$ where $n>1$ is an
integer. To discuss this, we first write the transverse metric
(for the case $N=8$) as
\be
\label{4-6} d\sigma^2 = S_2 T_2 U_2 |\eta(S) \eta(T) \eta(U)|^4
{\left| \prod_{i=1}^{8} (z-z_i)^{- 1 / 4} \right| } ^2 dz
d\bar{z}\ .
\ee
For the generic case where the locations $z_i$ of the strings are
different, we have a deficit angle of $\pi / 2$ for each string
and hence no orbifold singularities occur. However, when some of
the $z_i$ are identified, such singularities appear. For example,
consider the case where the eight $z_i$ coalesce into three points
$z_1$, $z_2$ and $z_3$, of orders three, three and two
respectively. Then the transverse metric turns to be
\be
\label{4-7} d\sigma^2 = S_2 T_2 U_2 |\eta(S) \eta(T) \eta(U)|^4
{\left| (z-z_1)^{- 3 / 4} (z-z_2)^{- 3 / 4} (z-z_3)^{- 1 / 2}
\right| } ^2 dz d\bar{z}\ ,
\ee
and one recognizes the deficit angles of $3\pi/2$, $3\pi/2$ and
$\pi$ around $z_1$, $z_2$ and $z_3$ respectively. The transverse
space is thus a $\bT^2/\bZ_4$ orbifold as we can see from $\delta$
for the $n=4$ value. Another example is obtained by taking the
eight $z_i$ to coalesce into four points $z_1,\ldots,z_4$, of
order two each. Then the transverse metric turns to
\be
\label{4-8} d\sigma^2 = S_2 T_2 U_2 |\eta(S) \eta(T) \eta(U)|^4
{\left| \prod_{i=1}^{4} (z-z_i)^{- 1 / 2} \right| } ^2 dz
d\bar{z}\ ,
\ee
and one recognizes a deficit angle of $\pi$ for each string. The
transverse space is thus a $\bT^2/\bZ_2$ orbifold.

We may now consider sting sources located at the points $z_i$,
$z_j$, $z_k$ where the scalar fields $S$, $T$, $U$ diverge with
energy-momentum tensors of the form
\be
\label{4-9} T_{\mu\nu}=-{\eta_{\mu\nu}} \left(\sum_{i=1}^{N_S}
{T_i} \delta^{(2)}(z-z_i) +\sum_{j=1}^{N_T} {T_j}
\delta^{(2)}(z-z_j)+\sum_{k=1}^{N_U} {T_k}
\delta^{(2)}(z-z_k)\right)\, , ~~~~~T_{z\bar{z}}=0\, .
\ee
In this case, the solution  \eqr{4-5} changes to
\ba
\label{4-10} d\sigma^2 &=& S_2 T_2 U_2 |\eta(S) \eta(T) \eta(U)|^4
\left| \prod_{i=1}^{N_S} \prod_{j=1}^{N_T} \prod_{k=1}^{N_U}
\left( (z-z_i) (z-z_j) (z-z_k) \right)^{- 1 / 12} \right|^2 \nn\\
&&~~~~~\times \left|\prod_{i=1}^{N_S} \prod_{j=1}^{N_T}
\prod_{k=1}^{N_U}  (z-z_i)^{-{T_i}/{2 \pi {M_*^{2}}}}
(z-z_j)^{-{T_j}/{2 \pi {M_*^{2}}}} (z-z_k)^{-{T_k}/{2 \pi
{M_*^{2}}}}\right|^2dz d\bar{z}\,
\ea
So at infinity the transverse space compactifies to $\bS^2$, if
\be
\label{4-11} 4\pi=\frac{\pi}{6} (N_S+N_T+N_U)
+\sum_{i=1}^{N_S}\frac{T_i}{M_*^{2}} +
\sum_{j=1}^{N_T}\frac{T_j}{M_*^{2}} +
\sum_{k=1}^{N_U}\frac{T_k}{M_*^{2}}\ .
\ee
Note that by imposing string/string/string triality we are led
again to take $N_S=N_T=N_U=N$ in \eqr{4-11}. Then again there are
no curvature singularities when $N_S=N_T=N_U=1$ and $T > 2\pi
M_{*}^{2}$.

\subsection{String solutions in $D=4$, $\cN=2$ supergravities}

We next consider the case of the $\cN=2$ theories arising from
compactifications of heterotic string theories on $K3 \times
\bT^2$ or, equivalently, from compactifications of minimal $\cN=1$
six-dimensional supergravity coupled to vector multiplets on
$\bT^2$ \cite{Duff:1995sm}. For these models, the moduli space
consists of (i) the vector-multiplet moduli space $\cM_V$
parametrized by the axion-dilaton $S$, the moduli $T$ and $U$
corresponding to combinations of the complex and K\"ahler
structure moduli $T^i$ of $\bT^2$, and the Wilson line moduli
$V^a$, and (ii) the hypermultiplet moduli space $\cM_H$
parametrized by the moduli of $K_3$ and of the vector bundle.
Restricting to the vector multiplet moduli space, its classical
geometry is locally of the form
\be
\label{4-12} \cM_V = \left( PSL(2,\bbZ) \setminus { SL(2,\bbR)
\over U(1)} \right) \times \left( SO(2,n;\bbZ) \setminus
{SO(2,n)\over {SO(2)\times SO(n)}} \right) \ ,
\ee
where $n=p+2$ with $p$ being the number of Wilson line moduli.
Here, the $PSL(2,\bbZ)$ and $SO(2,n;\bbZ)$ are the S- and
T-duality groups \cite{Giveon:1994fu}. This moduli space falls
into the class of special K\"ahler manifolds, considered in
section 3.

For the construction of solutions of interest in the models
considered here, there are two points that need special attention.
First, the $PSL(2,\bbZ)$ S-duality is no longer expected to be a
symmetry of the full quantum theory and so consistent solutions
can be constructed only by fixing the $S$ modulus to a constant
value and demanding invariance only under the T-duality group.
Second, the prepotential is renormalized both perturbatively and
nonperturbatively, where the $\cN=2$ non-renormalization theorems
guarantee that the perturbative corrections enter only at
one-loop order. Perturbatively exact solutions can thus be
constructed by taking account of the one-loop corrections which,
at the level of the K\"ahler potential, amount to the shift
\be
\label{4-13} S_2 \to S_2 + V_{GS}(T^i,V^a)\ ,
\ee
where $V_{GS}(T^i,V^a)$ is the Green-Schwarz term. Note that $S$
and $V_{GS}(T^i,V^a)$ transform under T-duality in such a way
that the corresponding transformation of $K$ is a K\"ahler
transformation. Given these observations, we may proceed to
construct stringy cosmic string solutions for the special cases
$n=1,2,3$ where the modular forms used for the construction of
invariant solutions are explicitly known.

\subsubsection{The $n=1$ $ST$ model}

The $ST$ model corresponds to the case where Wilson line moduli
are absent and only the $T$ modulus of the torus is turned on
\cite{Kaplunovsky:1995tm}. It is obtained from the general case
by setting $n=1$. The T-duality group is then
\be
\label{4-14} SO(2,1;\bbZ) \cong PSL(2,\bbZ) \ ,
\ee
and the classical K\"ahler potential is read off from Eq.
\eqr{3-12},\eqr{3-7}
\be
\label{4-15} K(S,T) = - \ln (S - \bar{S}) - 2 \ln (T - \bar{T})\ .
\ee
In the quantum theory, the above relation is modified by setting
$S_2 \to S_2 + V_{GS}(T)$.

As remarked earlier on, the stringy cosmic string solutions of
interest are constructed by fixing the $S$ modulus to some
constant value and imposing invariance under $PSL(2,\bbZ)_T$ and
non-degeneracy of the metric. The former requirement now leads to
a factor of $|\eta(T)|^8$ while the second requirement leads to a
factor of $|(z-z_i)^{-1/6}|^2$ for each string (the different
powers are due to the factor of two appearing in the K\"ahler
potential). Therefore, our solution for the transverse metric
reads
\be
\label{4-16} ds^2 = - d t^2 + d x^2 + (S_2 + V_{GS}) {T_2}^2
|\eta(T)|^8 {\left| \prod_{i=1}^{N} (z - z_i)^{- 1/6} \right|}^2
dz d\bar{z}\ .
\ee
At infinity for each string we have a deficit angle $\delta ={\pi
\over 3}$, and the total energy is $E={\pi \over 3} N$, i.e. the
energy per string is twice that of the $\cN=4$ solution. The
generalized solution becomes as \eqr{2-30}.

\subsubsection{The $n=2$ $STU$ model}

The $STU$ model corresponds to the case where Wilson line moduli
are absent and both moduli of the torus are turned on \cite{de
Wit:1995zg,Duff:1995sm}. It is obtained by the general case by
setting $n=2$. The classical T-duality group is in this case
\be
\label{4-17} SO(2,2;\bbZ) \cong PSL(2,\bbZ)_T \times
PSL(2,\bbZ)_U\ ,
\ee
and the classical K\"ahler potential is read off from Eq. \eqr{3-14},
\eqr{3-7}
\be
\label{4-18} K(S,T,U) = -\ln (S - \bar{S}) -\ln\left( (T -
\bar{T}) (U - \bar{U}) \right)\ .
\ee
There is also a $\bZ_2$ symmetry corresponding to the exchange $T
\leftrightarrow U$. In the quantum theory, Eq. \eqr{4-18} is similarly
modified by setting $S_2 \to S_2 + V_{GS}(T,U)$, while the $\bZ_2$
symmetry mentioned above is broken.

The stringy cosmic string solution is constructed as before, and
the result for the transverse metric is
\be
\label{4-19} d \sigma^2 = (S_2 + V_{GS}) T_2 U_2
|\eta(T)\eta(U)|^4 \left| \prod_{i=1}^{N_T}\prod_{j=1}^{N_U} (z -
z_i)^{- 1/12}(z - z_j)^{- 1/12}\right|^2 dz d\bar{z}\ .
\ee
As  $|z| \to \infty$ for each string we have a deficit angle
$\delta ={\pi \over 6}$, the total energy is $E={\pi \over 6} (N_T
+ N_U)$ and due to the fact that the $\bZ_2$ exchange symmetry is
broken, the numbers $N_T$ and $N_U$ may be different. Regularity
of the solution requires $N_T + N_U=24$. The generalized solution
becomes as in \eqr{4-10} with the factors corresponding to the
$N_S$ strings omitted.

\subsubsection{The $n=3$ $STUV$ model}

The final case we will consider here is the $STUV$ model
\cite{Lopes Cardoso:1996nc}, which corresponds to turning on a
single Wilson line modulus in addition to the two moduli of the
torus. It is obtained by the general case by setting $n=3$ so that
classical T-duality group is
\be
\label{4-20} SO(2,3;\bbZ) \cong Sp(4,\bbZ)\ .
\ee
In the genus two case the moduli space is the quotient of the
Siegel upper half space by the modular group $PSp(4,\bbZ)$, which
can be taken to be the fundamental domain $\cF_2$, and is
parametrized by the period matrix $\Omega$, which transforms
according to $\Omega \to (A \Omega+B) (C\Omega + D)^{-1}$. This
matrix is specified as
\be
\label{4-21} \Omega  = \left(
\begin{array}{cc} T & V
\\ V & U\end{array} \right)\ .
\ee
A Siegel modular form $F_w$ of weight $w$ is defined as a
holomorphic function of $\Omega$ that transforms as
\be
\label{4-22} F_w(\Omega)\to{(\det(C\Omega+D))}^w F_w(\Omega)\ .
\ee
Any such form admits a Laurent expansion in the parameters
$q=e^{2\pi {\rm i} T}$, $r=e^{2\pi {\rm i} V}$ and $s=e^{2\pi {\rm
i} U}$. The graded ring of Siegel modular forms is generated
\cite{Klingen,Freitag} by four forms of weight $4$, $6$, $10$ and
$12$, namely by the two Eisenstein series $\psi_4$ and $\psi_6$ and
the two cusp forms $\chi_{10}$ and $\chi_{12}$. In the
degeneration limit $\epsilon \to 0$, where $Sp(4,\bbZ)$
degenerates to $SL(2,\bbZ)\times SL(2,\bbZ)$, the genus two
surface can be constructed from two tori with modular parameters
$q_1=e^{2 \pi {\rm i} \tilde{T}}$ and $q_2=e^{2 \pi {\rm i}
\tilde{U}}$ \cite{Tuite:1999id,Mason:2006dk}. These two tori are
joined by excising a disk of radius $|\epsilon|$ from each torus
and making an appropriate identification of two annular regions
around the excised disk. In this limit the relations between the
parameters $T$,$U$,$V$ and $\tilde{T}$, $\tilde{U}$, $\epsilon$
are as follows \cite{Mason:2006dk}
\be
\label{4-23} T = \tilde{T} + \cO(\epsilon^2)\ \qquad U = \tilde{U}
+ \cO(\epsilon^2)\ \qquad V = - \epsilon + \cO(\epsilon^3)\ .
\ee
Turning now to the classical K\"ahler potential, this is read off
Eq. \eqr{3-16},\eqr{3-7}
\ba
\label{4-24} K(S,T,U,V) &=& - \ln(S -
\bar{S}) - \ln\left( (T - \bar{T})(U - \bar{U}) - (V - \bar{V})^2
\right) \nn\\ &=& - \ln(S - \bar{S}) - \ln\det (\Omega -
\bar{\Omega})\ ,
\ea
Again, in the quantum theory, Eq. \eqr{4-24} is modified by
setting $S_2 \to S_2 + V_{GS}(T,U,V)$.

The stringy cosmic string solution for the model under
consideration is obtained by generalizing the standard procedure
to the $Sp(4,\bbZ)$ case. First, the space of inequivalent choices
for $\Omega$ is, as said, the fundamental domain $\cF_2$ specified
by the conditions
\ba
\label{4-25} &&|T_1|,|U_1|,|V_1| \leqslant {1 \over 2}\ ,\qquad 0
\leqslant
|2V_2| \leqslant T_2 \leqslant U_2\ , \nn\\
&& |\det (C \Omega +
D) | \geqslant 1 \hbox{ for all } \left( \begin{array}{cc} A & B \\
C & D \end{array} \right) \in Sp(4,\bbZ)\ .
\ea
To construct finite-energy solutions, we need a set of holomorphic
functions that provide a map from the variable $\Omega$, which is
restricted to live on $\cF_2$ according to \eqr{4-25}, to the
Riemann sphere, i.e. the $Sp(4,\bbZ)$ counterparts of the
$J$-function. Such functions exist (known as Igusa invariants \cite{Igusa}) and are explicitly given in
terms of the $Sp(4,\bbZ)$ Eisenstein series $\psi_4$, $\psi_6$ and
the cusp forms $\chi_{10}$, $\chi_{12}$ as follows
\be
\label{4-26} x_1 = {\psi_4 \chi_{10}^2 \over \chi_{12}^2}\ ,\qquad
x_2 = {\psi_6 \chi_{10}^3 \over \chi_{12}^3}\ ,\qquad x_3 =
{\chi_{10}^6 \over \chi_{12}^5}\ .
\ee
Using the K\"ahler potential \eqr{4-24}, we find the transverse
metric
\be
\label{4-27} d \sigma^2 = (S_2 + V_{GS}) \det \Omega_2 |F(z)|^2 dz
d\bar{z}\ ,
\ee
where  $\Omega_2$ equals to $\Im \Omega$ and now the function
$F(z)$ must be chosen so as to enforce $Sp(4,\bbZ)$ modular
invariance and non-degeneracy of the metric. To ensure modular
invariance, we note that the $Sp(4,\bbZ)$ transformation of $\det
\Omega_2$ reads
\be
\label{4-28} \det \Omega_2 \to {\det \Omega_2 \over |\det(C\Omega
+ D)|^2}\
\ee
and hence can be compensated by multiplying by $|f(\Omega)|^2$,
where $f(\Omega)$ is an $Sp(4,\bbZ)$ modular form of weight $1$ as
follows from \eqr{4-22} with no zeros on the fundamental domain
$\cF_2$. The unique form with these properties is given by the
twelfth root of the cusp form $\chi_{12}$, i.e.
$f(\Omega)=\chi_{12}^{1/12}(\Omega)$. In order to have
non-degenerate metric, we note that the poles of Igusa invariants
are determined by the zeros of the cusp form $\chi_{12}$, as one
can see from Eq.\eqr{4-26}. This cusp form has zeros in the
z-plane at the locus $q=s=0$, where the locations of the $T-$,
$U-$ string cores $z_i$, $z_j$ exist. As we go around such a string,
 $\Omega$  should undergo an $Sp(4,\bbZ)$ transformation generated by the $Sp(4,\bbZ)$ matrices
\be
\mathrm{T}_i=\left(%
\begin{array}{cc}
  1_{2\times 2}& s_i \\
  0 &  1_{2\times 2} \\
\end{array}%
\right)\, ,
\ee
where
\be
 s_1=\left(\begin{array}{cc}
  1~& ~0 \\
  0~ &  ~0 \\
\end{array}%
\right), ~~s_2=\left(\begin{array}{cc}
  0~& ~0 \\
  0~ &  ~1 \\
\end{array}%
\right),~~s_3= \left(\begin{array}{cc}
  0~& ~1 \\
  1~ &  ~0 \\
\end{array}%
\right).
\ee
This leads to the $Sp(4,\bbZ)$ jumps  $\Omega \to \Omega + s_i$, or in terms of $T,U,V$,
\be
T\to T+1, ~~U\to U+1,  ~~V\to V+1\, .
\ee
 These monodromies and
holomorphicity require that near the core of the string, we will
have
\be
\label{4-29} T \sim {1 \over {2 \pi {\rm i}}} \ln(z - z_i)\ ,
\qquad U \sim {1 \over {2 \pi {\rm i}}} \ln(z - z_j)\ , \qquad V
\sim {1 \over {2 \pi {\rm i}}} \ln(z - z_k) \ ,
\ee
so  that
\be
\label{4-30} q = e^{2 \pi {\rm i} T} \sim (z-z_i)\ , \qquad s =
e^{2 \pi {\rm i} U} \sim (z-z_j)\ , \qquad r = e^{2 \pi {\rm i} V}
\sim (z-z_k)\ .
\ee
Note that, due to \eqr{4-25},  $V$ should degenerate together with $T$ and/or $U$, i.e, $z_k$ should coincide with $z_i$ and/or $z_j$.

Turning to $\chi_{12}$, its full expansion is given in
\cite{Tuite:1999id}, which to leading order reads
\be
\label{4-31}\chi_{12} = 96 q s + \dots \ .
\ee
Then, from \eqr{4-30} and \eqr{4-31} follows that the form of $F(z)$ in
the transverse metric is determined to be $F(z)=\chi_{12}^{1/12}
\prod_{i=1}^{N_T} \prod_{j=1}^{N_U} (z-z_i)^{-1/12}
(z-z_j)^{-1/12}$. This leads to the modular invariant,
non-degenerate solution
\be
\label{4-32}  d \sigma^2 = (S_2 + V_{GS}) \det \Omega_2
|\chi_{12}|^{1/6} \left| \prod_{i=1}^{N_T}\prod_{j=1}^{N_U} (z -
z_i)^{- 1/12} (z - z_j)^{- 1/12}\right|^2 dz d\bar{z}\ .
\ee
As $|z|\to \infty$ for each string the deficit angle is $\delta ={
{\pi} \over {6}}$ and the energy is indeed finite,
\be
\label{4-33} E = {{\pi} \over 6} (N_T + N_U) \ .
\ee
Regularity of the solution demands that $N_{T} + N_{U} = 24$.

In the degeneration limit $\epsilon \to 0$, $Sp(4,\bbZ)$
degenerates to $SL(2,\bbZ)\times SL(2,\bbZ)$. In this limit, the
Eisenstein series $\psi_4$, $\psi_6$ and the cusp forms
$\chi_{10}$, $\chi_{12}$ take the form
\ba
\label{4-34} && \psi_4 = E_4(q_1) E_4(q_2) + \cO(\epsilon^2)\ \nn\\
&& \psi_6 = E_6(q_1) E_6(q_2) + \cO(\epsilon^2)\ \nn\\&& \chi_{10}
= \epsilon^2 \Delta(q_1) \Delta(q_2) + \cO(\epsilon^4)\ \nn\\ &&
\chi_{12} = \Delta(q_1) \Delta(q_2) + \cO(\epsilon^2)\
\ea
where $E_4(q_i)$, $E_6(q_i)$, $\Delta(q_i)$, i=1,2 are the weight
$4$ and $6$ Eisenstein series and the cusp form of weight $12$
respectively for each $SL(2,\bbZ)$ factor. Then a linear
combination of Igusa invariants $x_1$,$x_2$,$x_3$, gives again a
modular invariant form. In particular, using the linear
combination
\be
\label{4-35} {{\alpha {(x_1)}^3 + \beta {(x_2)}^2 -\gamma x_3}
\over {x_3}} = {{\alpha {(\psi_4)}^3 + \beta {(\psi_6)}^2 -\gamma
\chi_{12}} \over {\chi_{12}}} \
\ee
where $\alpha ={41 \over 72}$, $\beta ={31 \over 72}$, $\gamma =
732096$ and substituting the expressions \eqr{4-34} for $\psi_4$,
$\psi_6$, $\chi_{12}$, which are valid in the limit $\epsilon \to
0$, leads to the modular invariant form $J(q_1) J(q_2)$
corresponding to the $SL(2,\bbZ)\times SL(2,\bbZ)$ case.

Now the zeros of the cusp form $\chi_{12}$ are easily found, as
for $\epsilon \to 0$ $\chi_{12} \to \Delta(q_1) \Delta(q_2)$, so
that $\chi_{12} \to 0$ for $(q_1,\epsilon) \to (0,0)$ and
$(q_2,\epsilon) \to (0,0)$. This implies that near the zeros $z_i$
and $z_j$, we can write $\det\Omega_2 |\chi_{12}^{1/12}|^2 \sim
|(z-z_i)^{1/12}(z-z_j)^{1/12}|^2$. Therefore, the appropriate
choice for $F(z)$ is $F(z)=\chi_{12}^{1/12}
\prod_{i=1}^{N_{\tilde{T}}} \prod_{j=1}^{N_{\tilde{U}}}
(z-z_i)^{-1/12} (z-z_j)^{-1/12}$. Then using the fact that
$\chi_{12} \to \Delta(q_1) \Delta(q_2)$ and $\det\Omega_2 = T_2
U_2$, one recovers the solution of the $STU$ model appeared in the
previous section.

\subsection{Brane solutions in $D=8$, $\cN=1$ supergravity}

Another situation where K\"ahler manifolds of the type $\cK_n =
{SO(2,n)\over {SO(2)\times SO(n)}}$ examined in section 3 occur is
$\cN=1$ supergravity coupled to $n$ vector multiplets in eight
dimensions \cite{Salam:1985ns},. Each vector multiplet contains 2 scalars so that the total  $2n$ scalars parametrize the
coset $\cK_n$. For this case, we can construct codimension-two solutions
for $n=1,2,3$ corresponding to five-branes. The
$n=1,2$ cases also appear as solutions to minimal $D=9$ and $D=7$
supergravities \cite{Gates:1984kr} coupled to two vector
multiplets.

Starting from codimension-two solutions, these can be constructed
by considering the ${SO(2,n)\over {SO(2)\times SO(n)}}$ K\"ahler
potential ($K_2$ in the notation of section 3). The resulting
transverse metrics are readily obtained from those of section 4.2
by simply discarding the $S$ modulus. Therefore, for the case
$n=1$ where there exists a single modulus $T$, we obtain the
solution
\be
\label{5-1} ds^2 = - d t^2 + d x_5^2 + {T_2}^2 |\eta(T)|^8 {\left|
\prod_{i=1}^{N} (z - z_i)^{- 1/6} \right|}^2 dz d\bar{z}\ ,
\ee
where we will denote by $dx_p^2$  the spatial metric on the world-volume of a p-brane.
The generalized solution is like \eqr{2-30}. In the $D=9$, $N=1$
supergravity coupled to $n$ vector multiplets the scalars
parametrize the coset $SO(1,n)/SO(n)$. It is clear that for two
vector multiplets coupled to gravity the codimension-two solution
is like \eqr{5-1}.

For the case $n=2$ where there exist two moduli $T$ and $U$, we
find
\be
\label{5-2} ds^2 = - d t^2 + d x_5^2 + T_2 U_2 |\eta(T)\eta(U)|^4
\left| \prod_{i=1}^{N_T}\prod_{j=1}^{N_U} (z - z_i)^{- 1/12}(z -
z_j)^{- 1/12}\right|^2 dz d\bar{z}\ .
\ee
Finally, for the case $n=3$ where there exist the three moduli
$T$, $U$ and $V$, we have
\be
\label{5-3} ds^2 = - d t^2 + d x_5^2 + \det \Omega_2
|\chi_{12}|^{1/6} \left| \prod_{i=1}^{N_T}\prod_{j=1}^{N_U} (z -
z_i)^{- 1/12} (z - z_j)^{- 1/12}\right|^2 dz d\bar{z}\ .
\ee
The generalized solution is as \eqr{4-10} by discarding the $N_S$
strings. In the $D=7$, $N=2$ supergravity coupled to $n$ vector
multiplets, $3n$ scalars parametrize the coset
$SO(3,n)/SO(3)\times SO(n)$. When the number of vector multiplets
is two then the codimension-two solution is like \eqr{5-3}.

Turning to codimension-four solutions, these can be constructed
according to guidelines of section 2.2. The simplest possible
situation is when the K\"ahler potential $K$ decomposes as in Eq.
\eqr{2-33} and is realized when $n=2$, in which case we have
\be
\label{5-4} K(T,U) = K^{(1)}(T) + K^{(2)}(U) = -\ln (T - \bar{T})
-\ln(U - \bar{U}) \ .
\ee
Then, setting $T=T(z)$ and $U=U(w)$ we obtain $k_{1\bar{1}}=T_2
|F^{(1)}(z)|^2$ and $k_{2\bar{2}}=U_2 |F^{(2)}(w)|^2$. Determining
the functions $F^{(1)}(z)$ and $F^{(2)}(w)$ in the usual manner,
we finally obtain the metric \eqr{2-36}, in the presence of
tensions
\ba
\label{5-5} ds^2 &=& - d t^2 + d x_3^2 + T_2 |\eta(T)|^4 \left|
\prod_{i=1}^{N_T} (z - z_i)^{- 1/12}\right|^2
\left|\prod_{i=1}^{N_T} (z-z_i)^{-{T^{(1)}_i}/{2 \pi
{M_*^{6}}}}\right|^2dz d\bar{z}\nn\\&&+ U_2 |\eta(U)|^4 \left|
\prod_{j=1}^{N_U}(w - w_j)^{- 1/12}\right|^2
\left|\prod_{j=1}^{N_U} (w-w_j)^{-{T^{(2)}_j}/{2 \pi
{M_*^{6}}}}\right|^2 d w d\bar{w}\ ,
\ea
The deficit angles at infinity in the $z$,$w$-plane are
\be
\label{5-6} \delta_1=\frac{\pi}{6}
N_T+\sum_{i=1}^{N_T}\frac{T^{(1)}_i}{M_*^{6}}\, , ~~~~
\delta_2=\frac{\pi}{6}
N_U+\sum_{j=1}^{N_U}\frac{T^{(2)}_j}{M_*^{6}}\, ,
\ee
respectively. With $\delta_i=4\pi$, we get an $\bS^2\times \bS^2$
compactification of the $D=8, \cN=1$ supergravity.

We may easily interpret the solution \eqr{5-5} by calculating the
corresponding energy momentum tensor $T_{MN}$. We may write
$T_{MN}=T^{\sigma}_{MN}+\sum_i^{N_T}T^{(i)}_{1,{MN}}+\sum_j^{N_U}T^{(j)}_{2,{MN}}$,
where $T^{\sigma}_{MN}$ is the scalar energy-momentum tensor and
$T^{(i)}_{MN}$ is the contribution of the brane located at the
point $z_i$. Then, by going to real coordinates
$z=x^4+ix^5,w=x^6+ix^7$ we find that
\ba \label{5-7}
&&T^{(i)}_{1, {\mu\nu}}=- {\eta_{\mu\nu}} T^{(1)}_i\delta^{(2)}(z-z_i)\, , ~~T^{(i)}_{1, {mn}}
=- {g_{mn}} T^{(1)}_i\delta^{(2)}(z-z_i)\, , ~~T^{(i)}_{1, {rs}}=0\, , \nn \\
&&T^{(j)}_{2, {\mu\nu}}=- {\eta_{\mu\nu}}
T^{(2)}_j\delta^{(2)}(w-w_j) \, , ~~T^{(j)}_{2, {mn}}=0\, ,
~~T^{(j)}_{2,{rs}}=- {g_{rs}} T^{(2)}_j\delta^{(2)}(w-w_j)\, .
\ea
where $(m,n=4,5)$ and  $(r,s=6,7)$ in \eqr{5-7} represents
intersecting five-branes of tensions $T^{(1)}$, $T^{(2)}$ with
world-volumes extended across the $(012345)$ and $(012367)$
directions. Their common world-volume in the $(0123)$ direction is
the 4D Minkowski intersection.

\section{Conclusions}

We have presented here  codimension-two
solutions of supergravity models in diverse dimensions, with or without brane sources.
We  have considered in particular $D$-dimensional
supergravity theories coupled to a set of scalar fields forming
 a nonlinear sigma model targeted on some
non-compact manifold. The scalar manifolds employed are  special K\"ahler  of the form
${SL(2,\bbR)\over U(1)} \times {SO(2,n) \over {SO(2)\times
SO(n)}}$ \cite{de Wit:1984pk,Ceresole:1995jg,Ferrara:1987tp} or the
Grassmannian cosets  ${SO(2,n) \over {SO(2)\times
SO(n)}}$ \cite{Salam:1985ns}.
The solutions we found are of the general form $M^{D-n}
\times \cK$, where $M^{D-n}$ is a flat Minkowski
space and $\cK$ is an $n$-dimensional internal space. We tried to keep the discussion as general as possible. However, for concreteness
we have  considered the cases of $\cN=4$ and $\cN=2$
supergravity in 4 dimensions  as well as  minimal supergravities coupled
to vector multiplets in 8 dimensions. In the first case, the solution  presents a string and
the 4D spacetime is compactified down to two-dimensions by a number of such strings. In the former case, the solution presents a five-brane or intersecting five-branes
along four-dimensional flat space, compactifying this way the eight-dimensional supergravity down to 4D Minkowski space-time.

The explicit solutions were found by employing a
holomorphic ansatz for the scalars. The latter were restricted to
lie in the fundamental domain of the modular groups
and allowing modular $SL(2,\bbZ), ~SL(2,\bbZ)\times SL(2,\bbZ)$ or $Sp(4,\bbZ)$  jumps around certain points in the internal space.
This modular jumps permit  scalar field configurations with finite
energy per unit volume and  explicit solutions
presented only  for those cases
where the modular forms required to construct the solutions were
explicitly known. Note that in order the solutions to have finite energy, one has to
arrange the total deficit angle produced by  the scalar configurations  to be
$4 \pi$ in which case the internal space  compactifies to $\bS^2$.

All the  solutions we have described have  be  generalized to include  brane
configurations as well, with the only requirement that  the scalars of the theory do not couple to the branes.
 The latter induces further deficit angles,
proportional to their tensions in the internal space. The requirement for the absence of conical
singularities may be fulfilled by suitably
tuning the
brane tensions  so that the total deficit angle equals $4 \pi$ and leading to a smooth sphere compactification.
It should also be noted that configurations of this type
might be relevant for the solution of the cosmological constant problem as the world-volume of the branes are
always flat irrespectively of any bulk dynamics.

\vskip.3in \noindent {\bf Acknowledgment} This work is co - funded
by the European Social Fund (75\%) and National Resources (25\%) -
(EPEAEK II) -PYTHAGORAS. We would like to thank S. Randjbar-Daemi
for extensive correspondences and detailed comments on the paper.

\end{document}